\documentclass{article}
\usepackage{waspaa19,amsmath,graphicx,url,times}
\usepackage[capitalize]{cleveref}
\usepackage{color}

\title{Speech Bandwidth Extension With WaveNet}

\name{Archit Gupta, Brendan Shillingford, Yannis Assael, Thomas C. Walters}
\address{DeepMind}

\begin{document}

\ninept
\maketitle

\begin{sloppy}

\begin{abstract}
Large-scale mobile communication systems tend to contain legacy transmission channels with narrowband bottlenecks, resulting in characteristic `telephone-quality' audio.
While higher quality codecs exist, due to the scale and heterogeneity of the networks, transmitting higher sample rate audio with modern high-quality audio codecs can be difficult in practice.
This paper proposes an approach where a communication node can instead extend the bandwidth of a band-limited incoming speech signal that may have been passed through a low-rate codec.
To this end, we propose a WaveNet-based model conditioned on a log-mel spectrogram representation of a bandwidth-constrained speech audio signal of 8 kHz and audio with artifacts from GSM full-rate (FR) compression to reconstruct the higher-resolution signal.
In our experimental MUSHRA evaluation, we show that a model trained to upsample to 24kHz speech signals from audio passed through the 8kHz GSM-FR codec is able to reconstruct audio only slightly lower in quality to that of the Adaptive Multi-Rate Wideband audio codec (AMR-WB) codec at 16kHz, and closes around half the gap in perceptual quality between the original encoded signal and the original speech sampled at 24kHz.
We further show that when the same model is passed 8kHz audio that has not been compressed, is able to again reconstruct audio of slightly better quality than 16kHz AMR-WB, in the same MUSHRA evaluation.

\end{abstract}


\section{Introduction and Related Work}
\label{sec:intro}

Legacy transmission channels are still part of many large-scale communication systems.
These channels introduce bottlenecks, limiting the bandwidth and the quality of speech. Often this is referenced as `telephone-quality' audio.
Upgrading all parts of the infrastructure to be compatible with higher quality audio codecs  can be difficult.
Thus, instead of upgrading all the communication nodes of the infrastructure, this paper proposes an approach where a communication node can instead extend the bandwidth of any incoming speech signal. To achieve that we propose a model based on WaveNet \cite{van2016wavenet}, a deep generative model of audio waveforms.

\begin{figure}[!t]
  \centering
  \centerline{\includegraphics[width=\columnwidth]{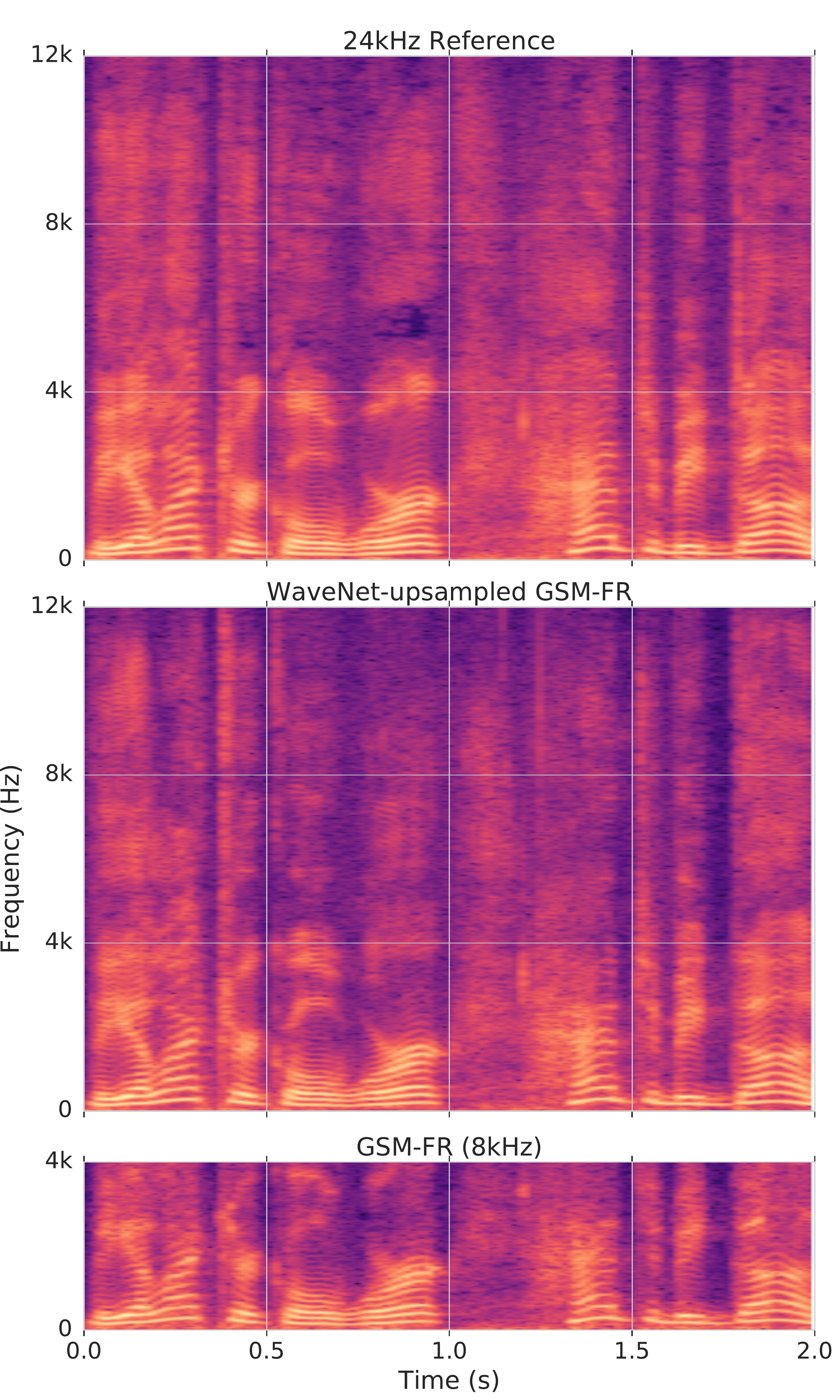}}
  \caption{Spectrograms from an utterance from the LibriTTS corpus. Top: Original audio, Middle: Audio reconstructed from the WaveNet model conditioned on spectrograms derived from GSM-FR audio, Bottom: Spectrogram from GSM-FR audio.}
  \label{fig:spectrograms}
\end{figure}

WaveNet has been shown to be extremely effective at synthesizing high quality speech when conditioned on linguistic features. In addition, the WaveNet architecture has been used conditioned on log-mel spectrograms for text-to-speech \cite{tacotron2} and other low-dimensional latent representations for speech coding \cite{chrome,vqvaeaudiocoding}. 
Given the power of the WaveNet architecture to generate high quality speech audio from a constrained conditioning representation, we extend this technology is to the problem of bandwidth extension (BWE) \cite{bwebook} for speech, also known as audio super-resolution \cite{audiosuperres}.

While BWE can be taken to mean the extension of a band-limited signal to both lower- and higher-frequency regions, in this case we are particularly interested in the application to telephony, where audio is often passed through a low-rate speech codec like GSM full-rate (FR) \cite{gsm06.10} which limits the highest frequency components of the reconstructed signal to be under 4kHz, leading to lower audio quality and potential impairment of intelligibility. Therefore we focus on the reconstruction of signals with a sample rates of 24kHz from input signals with a sample rate of 8kHz.
Historically, bandwidth extension has been performed in the domain of vocoder representations of speech, using techniques like Gaussian mixture models and hidden Markov models \cite{bwebook}; more recently interest has grown in using neural networks either to model spectral envelopes \cite{bwespectral} or to predict the upsampled waveform directly \cite{audiosuperres,hrnn,dilatedconvbwe}, leading to quality gains over earlier methods.

In our experimental evaluation, we assess the ability of our proposed model to perform bandwidth extension on narrowband signals. To illustrate the impact of our work, we show that a model trained to upsample to 24kHz speech signals passed through a GSM-FR codec at 8kHz is able to reconstruct audio which is of similar or better quality to that produced by the Adaptive Multi-Rate Wideband (AMR-WB) \cite{3gpp.26.071} codec at 16kHz. GSM-FR is a codec used in legacy GSM mobile phone calls, while AMR-WB is the codec commonly used for `HD-voice' calls. Although it is difficult to compare with previous work, because of lack of reproducible code and different test set splits our method achieves higher scores in a MUSHRA evaluation compared to previous work \cite{audiosuperres}.

It is worth mentioning that we believe our WaveNet core could potentially be replaced with more computationally efficient architectures such as parallel WaveNet \cite{parallelwavenet}, WaveGlow \cite{waveglow} or WaveRNN \cite{wavernn}. These architectures have shown that it is often possible to reproduce a more computationally-tractable version of a model while maintaining similar modelling performance. In this work, we build a proof-of-concept for high-quality bandwidth extension built on WaveNet due to its superior representational power and relative ease of training, leaving open the possibility of reproducing the results using other, more computationally tractable, architectures.

\section{Training Setup}
\subsection{Model Architecture}
WaveNet is a generative model that models the joint probability of a waveform $\mathbf{x}=\left\{x_{1}, \dots, x_{T}\right\}$ as a product of conditional probabilities given the samples at previous timesteps. A conditional WaveNet model takes an additional input variable $\mathbf{h}$ and models this conditional distribution as
$$
p(\mathbf{x} | \mathbf{h})=\prod_{t=1}^{T} p\left(x_{t} | x_{1}, \ldots, x_{t-1}, \mathbf{h}\right).
$$
A conditional WaveNet model is used in this task. The conditioning inputs $\mathbf{h}$ are passed through a `conditioning stack' consisting of five dilated convolution layers, followed by two transpose convolutions, which have the effect of upsampling the conditioning input by a factor of four.
Autoregressive inputs are normalized in the range [-1, 1] and passed through a convolutional layer with filter size $4$ and $512$ filters. They are then fed into the core WaveNet model, which consists of three stacks of $10$ dilated convolution layers, with skip connections, as in the original WaveNet architecture \cite{van2016wavenet}. The dilation factor we use is 2; the filter size and number of filters are 3 and 512 respectively. The output from the skip connections is passed through two convolutional layers with $256$ filters each.
The output distribution over sample values is modelled using a quantized logistic mixture \cite{pixelcnnpp}, with 10 components.

\subsection{Data preparation}
Our models were trained and evaluated on the LibriTTS \cite{zen2019libritts} dataset.
LibriTTS
is derived from the same source materials as the well-known LibriSpeech corpus \cite{panayotov2015librispeech}, but contains audio sampled at 24kHz (as opposed to 16kHz for LibriSpeech), with a resolution of 16 bits per sample. Both datasets are derived from a collection of public-domain audiobooks (and associated text) read by English speakers with a range of accents and in a variety of non-studio conditions, meaning that there is often some background noise in the recordings. The \texttt{train-clean-100} and \texttt{train-clean-360} subsets of the data were variously used for training, with a small proportion (1-2\%) of each set held out for evaluation.
Listening evaluations were conducted on the \texttt{test-clean} subset, which contains a disjoint set of speakers from the train sets, ensuring that no speakers present in the train set were used.

\subsection{Training}

The model is trained using maximum likelihood to predict the 24kHz waveform from log mel-spectrograms computed from the 8kHz band-limited waveform.
As with other instances of WaveNet, there were two types of inputs to the model during training, the autoregressive inputs containing the sample from the previous timestep, and the conditioning inputs. The autoregressive inputs during training are teacher-forced, and were therefore fed high quality 24kHz audio samples. 
We compute log-mel spectrograms from the lower bandwidth audio as conditioning inputs.

In other words, the WaveNet described previously then models:
$$
p(\mathbf{x}_{\text{hi}} | \mathbf{x}_{\text{lo}})=\prod_{t=1}^{T} p\left(x_{\text{hi},t} | x_{\text{hi},1}, \ldots, x_{\text{hi},t-1}, \mathbf{x}_{\text{lo}}\right).
$$
where $\mathbf{x}_{\text{hi}}$ is the autoregressively modelled 24kHz waveform, and $\mathbf{x}_{\text{lo}}$ is the 8kHz band-limited version, represented as a log mel-spectrogram. The $\mathbf{x}_{\text{lo}}$ is used as input in the WaveNet conditioning stack.

We use the Adam \cite{kingma2015adam} optimizer with a learning rate of $10^{-4}$, momentum set to $0.9$, and epsilon set to $10^{-8}$. We use a total batch size of $64$ over $8$ Tensor Processing Unit (TPU) cores \cite{jouppi2017datacenter}, for a batch size of 8 per core.

\begin{figure}[t]
  \centering
  \centerline{\includegraphics[width=0.75\columnwidth]{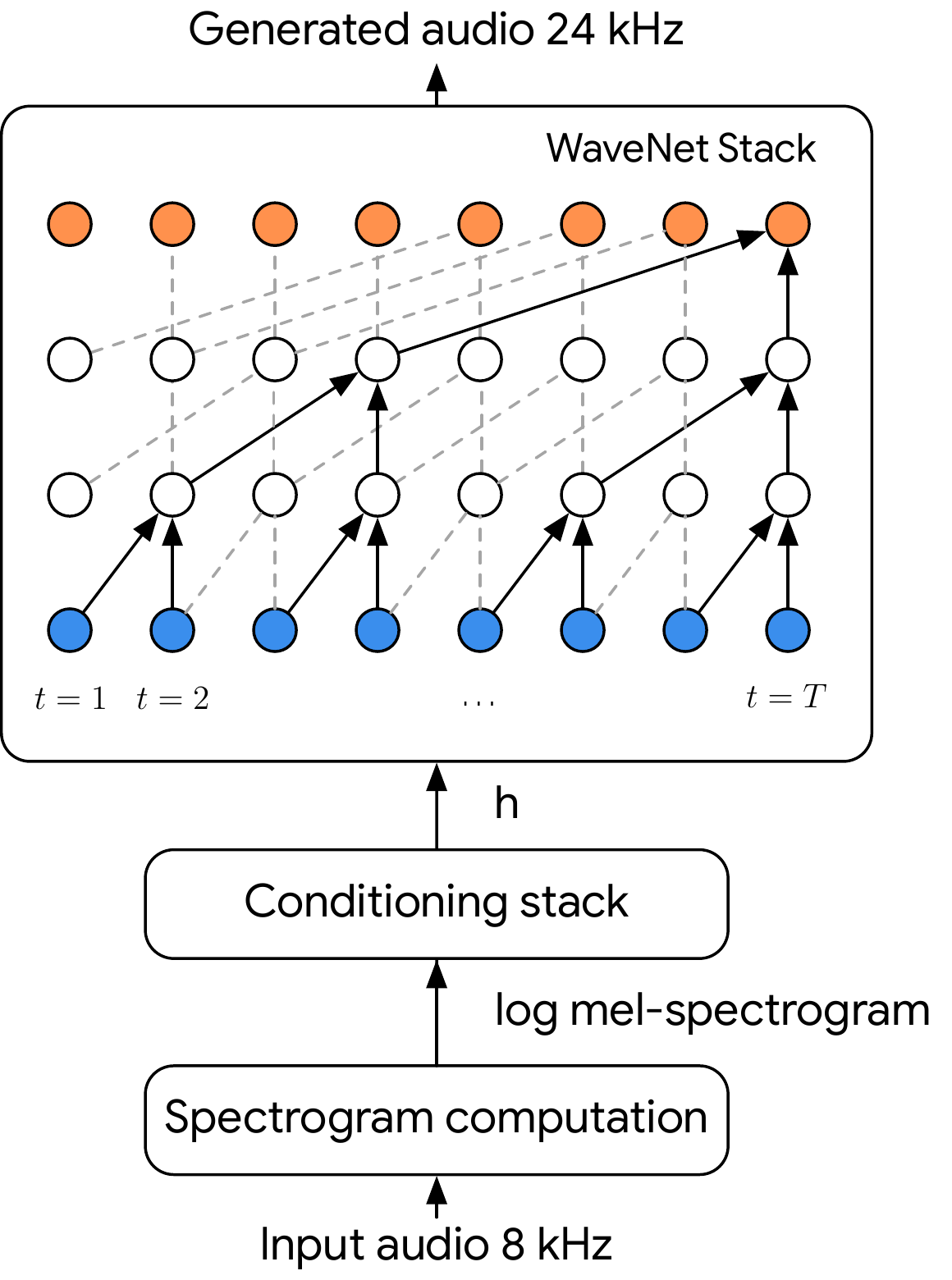}}
  \caption{Illustration of the processing pipeline. The input audio, sampled at 8 kHz, is transformed to a log mel-spectrogram representation, then used as input in the conditioning stack of WaveNet. The model outputs high-sample rate 24 kHz audio with higher frequencies predicted from the rest of the signal.}
  \label{fig:architecture}
\end{figure}

\section{Experimental evaluation}

\subsection{Setup}

In this evaluation we are primarily interested in the case of speech enhancement in the setting of a fixed legacy audio coding pathway, such as calls on standard GSM mobile networks. In this case, the codec typically operates with a bandwidth of 4kHz, leading to an audio waveform with an 8kHz sample rate.

To generate the training set, the LibriTTS clean-100 train set was preprocessed with the sox\footnote{\url{http://sox.sourceforge.net/}} tool, passing the original audio through the GSM-FR encoder, leading to a dataset containing an original 24kHz audio signal and a signal with an 8kHz sample rate and some further quality degradation from the application of the codec, for each utterance.
To generate a training pair given an utterance in the LibriTTS training set, a 350 ms audio region from a random point in the utterance was selected. Log-mel spectrograms were generated from the 8kHz input audio in the training region using a Hann window of 50ms with a step size of 12.5ms, then mapped to 80 mel-frequency bins ranging from 125Hz up to the Nyquist frequency of the input signal. These parameters lead to conditioning vectors $\mathbf{x}_\text{lo}$ of length 80 at a rate of 80Hz.
A WaveNet network was then trained to predict the ground-truth 24kHz sample-rate audio for the same region, given the spectrograms computed from the GSM audio as conditioning. In early experiments we found that this spectrogram-conditioned approach performed better compared to feeding in a raw waveform directly as conditioning.

\subsection{Results}
We evaluate our models using the MUltiple Stimuli with Hidden Reference and Anchor (MUSHRA) \cite{mushra} listening test methodology. Each listener is presented with the ground-truth 24kHz reference labelled as such, as well as several test items presented without being labelled: the 24kHz reference, AMR-WB encoded audio, GSM-FR encoded audio (the low-quality anchor), 8kHz audio (downsampled using the default settings in sox), the WaveNet-upsampled 8kHz-to-24kHz prediction and the WaveNet-upsampled GSM-FR-to-24kHz prediction. 

Raters are requested to give each test utterance a score between 0 and 100 using a slider with equally-spaced regions labelled `bad', `poor', `fair', `good' and `excellent'. Raters should score the hidden reference at close to 100, and the `anchor' stimulus should receive the lowest score. Typically MUSHRA evaluations are performed with a small set of trained raters. However, the raters used in this evaluation were untrained and so each utterance was rated by 100 different raters to ensure narrow error bars.

MUSHRA tests show that the model trained to predict to 24kHz from 8kHz audio directly performs slightly better than the AMR-WB codec, while the model predicting 24kHz from GSM encoded 8kHz performs only slightly worse than AMR-WB.

A set of samples was selected for the listening test from the LibriTTS \texttt{test-clean} corpus. Samples were chosen by randomly picking one utterance of between 3 and 4 seconds for each speaker in the test set. This led to 36 utterances of which 8 were selected at random for evaluation in the MUSHRA listening test. 

The MUSHRA listening test results are shown in \cref{fig:results}.

Finally, to visually illustrate the quality of a reconstructed sample, \cref{fig:spectrograms} depicts spectrograms of the original, the reconstructed, and the GSM-FR audio from an utterance from the LibriTTS corpus.

\begin{figure}[t]
  \centering
  \centerline{\includegraphics[width=\columnwidth]{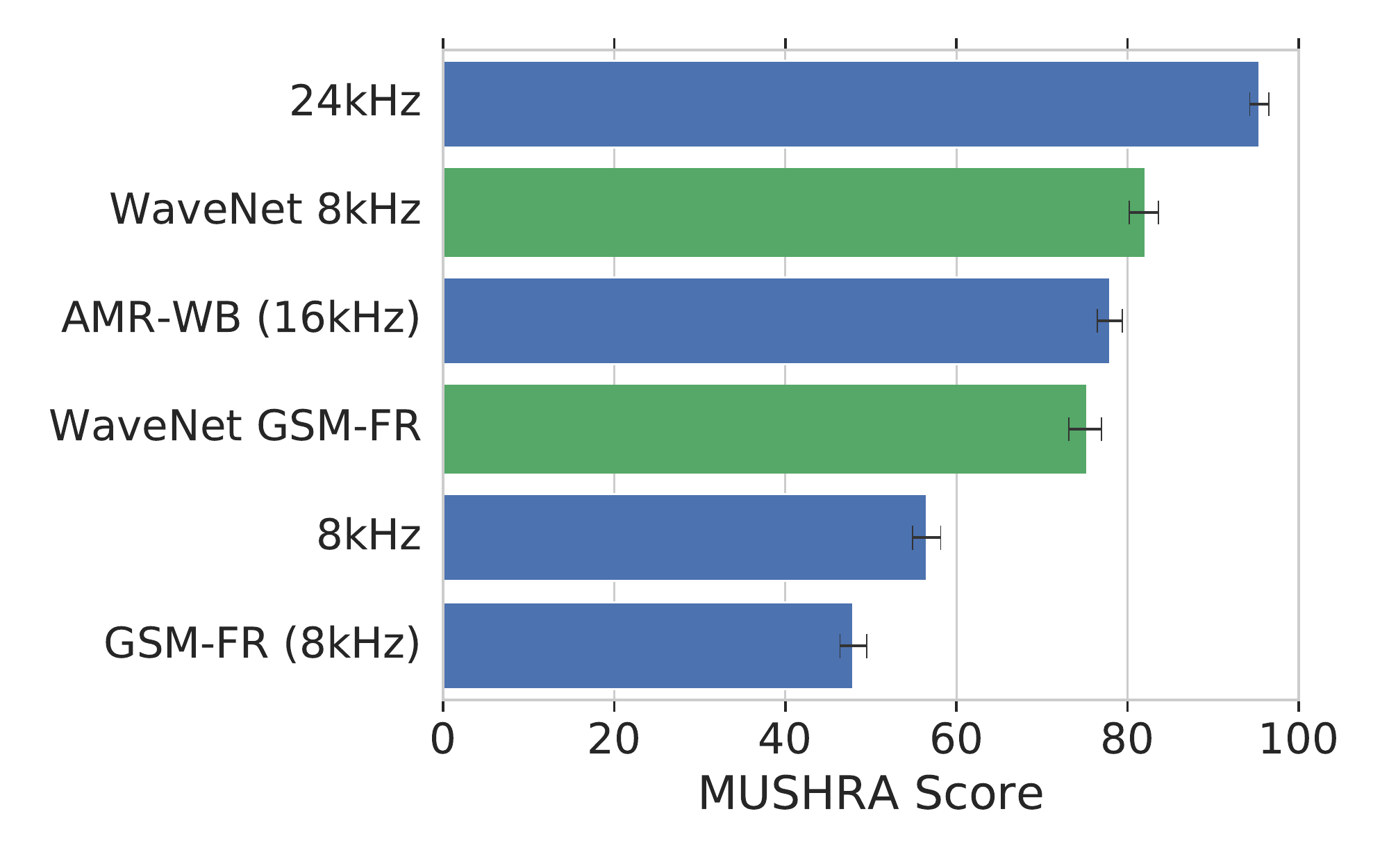}}
  \caption{Our model (\textsc{WaveNet 8kHz} and \textsc{WaveNet GSM-FR}), trained on 8kHz GSM-FR audio signals and evaluated with 8kHz uncompressed and 8kHz GSM-FR audio, is evaluated using the MUSHRA listening test methodology. The model is compared against the initial audio at \textsc{24kHz} and \textsc{8kHz}, and the \textsc{AMR-WB} 16kHz and \textsc{GSM-FR} 8kHz codecs.}
  \label{fig:results}
\end{figure}

\section{Conclusions}

This work introduces a new WaveNet-based model for speech bandwidth extension. The model is able to reconstruct 24kHz audio from 8kHz signals that is of similar or better quality to that produced by the AMR-WB codec at 16kHz. Our upsampling method produces ``HD-Voice''-quality audio from standard telephony-quality and GSM-quality audio, showing promise that our approach for audio super resolution is feasible for improving audio quality in existing telephony systems.
For future work, other architectures, such as WaveRNN, can be evaluated on the same task to improve computational efficiency.

\bibliographystyle{IEEEtran}
\bibliography{main}

\begin{thebibliography}{10}
\providecommand{\url}[1]{#1}
\def\UrlFont{\rmfamily}
\providecommand{\newblock}{\relax}
\providecommand{\bibinfo}[2]{#2}
\providecommand\BIBentrySTDinterwordspacing{\spaceskip=0pt\relax}
\providecommand\BIBentryALTinterwordstretchfactor{4}
\providecommand\BIBentryALTinterwordspacing{\spaceskip=\fontdimen2\font plus
\BIBentryALTinterwordstretchfactor\fontdimen3\font minus
  \fontdimen4\font\relax}
\providecommand\BIBforeignlanguage[2]{{%
\expandafter\ifx\csname l@#1\endcsname\relax
\typeout{** WARNING: IEEEtran.bst: No hyphenation pattern has been}%
\typeout{** loaded for the language `#1'. Using the pattern for}%
\typeout{** the default language instead.}%
\else
\language=\csname l@#1\endcsname
\fi
#2}}

\bibitem{van2016wavenet}
A.~{v}an~{d}en Oord, S.~Dieleman, H.~Zen, K.~Simonyan, O.~Vinyals, A.~Graves,
  N.~Kalchbrenner, A.~W. Senior, and K.~Kavukcuoglu, ``Wave{N}et: A generative
  model for raw audio.'' in \emph{SSW}, 2016, p. 125.

\bibitem{tacotron2}
J.~Shen, R.~Pang, R.~J. Weiss, M.~Schuster, N.~Jaitly, Z.~Yang, Z.~Chen,
  Y.~Zhang, Y.~Wang, R.~Skerrv-Ryan, \emph{et~al.}, ``Natural tts synthesis by
  conditioning wavenet on mel spectrogram predictions,'' in \emph{IEEE
  International Conference on Acoustics, Speech and Signal Processing
  (ICASSP)}.\hskip 1em plus 0.5em minus 0.4em\relax IEEE, 2018, pp. 4779--4783.

\bibitem{chrome}
W.~B. Kleijn, F.~S. Lim, A.~Luebs, J.~Skoglund, F.~Stimberg, Q.~Wang, and T.~C.
  Walters, ``Wave{N}et based low rate speech coding,'' in \emph{IEEE
  International Conference on Acoustics, Speech and Signal Processing
  (ICASSP)}.\hskip 1em plus 0.5em minus 0.4em\relax IEEE, 2018, pp. 676--680.

\bibitem{vqvaeaudiocoding}
C.~Garbacea, A.~van~den Oord, Y.~Li, F.~S.~C. Lim, A.~Luebs, O.~Vinyals, and
  T.~C. Walters, ``Low bit-rate speech coding with {VQ}-{VAE} and a {W}ave{N}et
  decoder,'' in \emph{IEEE International Conference on Acoustics, Speech and
  Signal Processing (ICASSP)}.\hskip 1em plus 0.5em minus 0.4em\relax IEEE,
  2019.

\bibitem{bwebook}
E.~R. Larsen and R.~M. Aarts, \emph{Audio Bandwidth Extension: Application of
  Psychoacoustics, Signal Processing and Loudspeaker Design}.\hskip 1em plus
  0.5em minus 0.4em\relax USA: John Wiley \&; Sons, Inc., 2004.

\bibitem{audiosuperres}
V.~Kuleshov, S.~Z. Enam, and S.~Ermon, ``Audio super resolution using neural
  networks,'' \emph{arXiv preprint arXiv:1708.00853}, 2017.

\bibitem{gsm06.10}
\BIBentryALTinterwordspacing
ESTI, ``{GSM Full Rate Speech Transcoding},'' {European Digital Cellular
  Telecommunications System}, Tech. Rep. 06.10, 02 1992, version 3.2.0.
  [Online]. Available:
  \url{https://www.etsi.org/deliver/etsi_gts/06/0610/03.02.00_60/gsmts_0610sv030200p.pdf}
\BIBentrySTDinterwordspacing

\bibitem{bwespectral}
J.~Abel and T.~Fingscheidt, ``Artificial speech bandwidth extension using deep
  neural networks for wideband spectral envelope estimation,'' \emph{IEEE/ACM
  Transactions on Audio, Speech, and Language Processing}, vol.~PP, pp. 1--1,
  10 2017.

\bibitem{hrnn}
Z.-H. Ling, Y.~Ai, Y.~Gu, and L.-R. Dai, ``Waveform modeling and generation
  using hierarchical recurrent neural networks for speech bandwidth
  extension,'' \emph{IEEE/ACM Transactions on Audio, Speech, and Language
  Processing}, vol.~26, no.~5, pp. 883--894, 2018.

\bibitem{dilatedconvbwe}
Y.~Gu and Z.-H. Ling, ``Waveform modeling using stacked dilated convolutional
  neural networks for speech bandwidth extension.'' in \emph{INTERSPEECH},
  2017, pp. 1123--1127.

\bibitem{3gpp.26.071}
\BIBentryALTinterwordspacing
3GPP, ``{Mandatory speech CODEC speech processing functions; AMR speech CODEC;
  General description},'' {3rd Generation Partnership Project (3GPP)},
  Technical Specification (TS) 26.071, 06 2018, version 15.0.0. [Online].
  Available:
  \url{https://portal.3gpp.org/desktopmodules/Specifications/SpecificationDetails.aspx?specificationId=1386}
\BIBentrySTDinterwordspacing

\bibitem{parallelwavenet}
A.~van~den Oord, Y.~Li, I.~Babuschkin, K.~Simonyan, O.~Vinyals, K.~Kavukcuoglu,
  G.~van~den Driessche, E.~Lockhart, L.~Cobo, F.~Stimberg, N.~Casagrande,
  D.~Grewe, S.~Noury, S.~Dieleman, E.~Elsen, N.~Kalchbrenner, H.~Zen,
  A.~Graves, H.~King, T.~Walters, D.~Belov, and D.~Hassabis, ``Parallel
  {W}ave{N}et: Fast high-fidelity speech synthesis,'' in \emph{Proceedings of
  the 35th International Conference on Machine Learning}, ser. Machine Learning
  Research, vol.~80.\hskip 1em plus 0.5em minus 0.4em\relax Stockholmsmässan,
  Stockholm Sweden: PMLR, 2018, pp. 3918--3926.

\bibitem{waveglow}
R.~Prenger, R.~Valle, and B.~Catanzaro, ``Waveglow: {A} flow-based generative
  network for speech synthesis,'' in \emph{IEEE International Conference on
  Acoustics, Speech and Signal Processing (ICASSP)}.\hskip 1em plus 0.5em minus
  0.4em\relax IEEE, 2019.

\bibitem{wavernn}
N.~Kalchbrenner, E.~Elsen, K.~Simonyan, S.~Noury, N.~Casagrande, E.~Lockhart,
  F.~Stimberg, A.~Oord, S.~Dieleman, and K.~Kavukcuoglu, ``Efficient neural
  audio synthesis,'' in \emph{International Conference on Machine Learning},
  2018, pp. 2415--2424.

\bibitem{pixelcnnpp}
T.~Salimans, A.~Karpathy, X.~Chen, and D.~P. Kingma, ``Pixelcnn++: A pixelcnn
  implementation with discretized logistic mixture likelihood and other
  modifications,'' in \emph{International Conference on Learning
  Representations (ICLR)}, 2017.

\bibitem{zen2019libritts}
H.~Zen, V.~Dang, R.~Clark, Y.~Zhang, R.~J. Weiss, Y.~Jia, Z.~Chen, and Y.~Wu,
  ``{LibriTTS}: A corpus derived from librispeech for text-to-speech,''
  \emph{arXiv preprint arXiv:1904.02882}, 2019.

\bibitem{panayotov2015librispeech}
V.~Panayotov, G.~Chen, D.~Povey, and S.~Khudanpur, ``Librispeech: an asr corpus
  based on public domain audio books,'' in \emph{IEEE International Conference
  on Acoustics, Speech and Signal Processing (ICASSP)}.\hskip 1em plus 0.5em
  minus 0.4em\relax IEEE, 2015, pp. 5206--5210.

\bibitem{kingma2015adam}
D.~P. Kingma and J.~Ba, ``{ADAM}: A method for stochastic optimization,'' in
  \emph{International Conference on Learning Representations (ICLR)}, 2015.

\bibitem{jouppi2017datacenter}
N.~P. Jouppi, C.~Young, N.~Patil, D.~Patterson, G.~Agrawal, R.~Bajwa, S.~Bates,
  S.~Bhatia, N.~Boden, A.~Borchers, \emph{et~al.}, ``In-datacenter performance
  analysis of a tensor processing unit,'' in \emph{International Symposium on
  Computer Architecture (ISCA)}.\hskip 1em plus 0.5em minus 0.4em\relax IEEE,
  2017, pp. 1--12.

\bibitem{mushra}
{International Telecommunication Union}, ``Method for the subjective assessment
  of intermediate sound quality ({MUSHRA}),'' ITU-R Recommendation BS.1534-1,
  Tech. Rep., 2001.

\end{thebibliography}

\end{sloppy}
\end{document}